# RECYCLER CHROMATICITIES AND END SHIMS FOR NOVA AT FERMILAB*

M. Xiao, Fermilab, Batavia, IL 60540, USA

*Abstract*

In era of NOvA operation, it is planned to slip-stack six on six Booster proton batches in the Recycler ring for a total intensity of $5\times10^{13}$ protons/cycle. During the slip-stacking, the chromaticities are required to be jumped from (-2,-2) to (-20,-20). However, with the existing 2 families of powered sextupoles in the lattice, the chromaticities can only be adjusted to (-12,-12) from (-2,-2). On the other hand, the presently designed Recycler lattice for Nova replaces the 30 straight section with 8 "D-D half FODO cells". With the limit of the feasible quad strength, 3 quads in a half-cell were used to obtain the working point under, and the maximum beta-functions in this section cannot be less than 80 m. In this paper, we re-designed the end shims of the permanent magnets in the ring lattice with appropriate quadrupole and sextupole components to meet both chromaticity and tune requirements. We are able to use 2 quads in a half cell in RR30 straight section within feasible quad strength. The maximum beta-functions are also lowered to around 55 m. The dynamic aperture tracking has been done using MAD to simulate the scenario of beam injection into the Recycler ring for Nova.

## INTRODUCTION

The Recycler Ring (RR) is comprised of 344 permanent gradient dipoles and 94 permanent quadrupoles. A phase trombone was installed in the RR-60 long straight section, which contains 9 families of trim quads, to locally adjust the phase within the straight without perturbing the lattice outside the insert. The gradient magnet in the Recycler Ring is a straight (rectangular) dipole, with uniform dipole, quadrupole and sextupole components along the length. The magnets were designed with end shims which correct the field shape (harmonic components up to 10-pole). End shims (small ~1 inch section of pole-tip) is used at each end of the gradient dipole to satisfy the field strength and uniformity specifications. The designed working point tunes are $Q_x$=25.425, $Q_y$=24.415, the chromaticities are $Q_x$'=-2, $Q_y$'=-2. In era of NOvA operation, it is planned to slip-stack six on six Booster proton batches in the Recycler ring for a total intensity of $5\times10^{13}$ protons/cycle. It required that the chromaticities to jump from (-2,-2) to (-20,-20), and additional tune compensations due to space charge tune shift of the high intensity beam during the slip-stacking.

We replace the RR-30 straight section with 8 "D-D half FODO cells", and install another trombone insert similar to the one in RR-60 to increase capability to adjust tunes. There are the 2 families of powered sextupoles in the existing RR lattice, but the chromaticities can only be adjusted to (-12,-12) from (-2,-2). We will design new end shims of the ARC gradient magnet to get corrected chromaticities of (-10,-10).

We use Fermilab unit $b_n$ to represent the strength of each field component $B_n$. The definition is;

$$b_n \times 10^{-4} = B_n r^n / B_o \ @\ r = 1"  \quad (1)$$

where $B_o$ is the guided field, the field strength of the dipole magnet.

## CHROMATICITIES OF (-10, -10)

It was found that the nature chromaticities without body sextupoles in the gradient dipole magnets are (-33,-34). The chromaticities with measured body sextupoles in the gradient dipole magnets are (-29, +18). The corrected chromaticities with the sextupoles in the present end shims are (-2,-2). There are types of end-shims [1]: (1) "Standard end-shim" at one end (upstream): identical for all the gradient magnets, which was chosen to correct for one-half of the average systematic defect as measured in the first ~20 magnets of the production run; (2) "Customized end shims" at the other end (downstream): Numerically controlled program generate the shape according to measured field of the magnet to cure the field errors.

Only "Standard end-shim" is going to be re-designed. Firstly, we obtained the sextupole component $(b_2)_F$ = 0.2318 and $(b_2)_D$=-3.3985 respectively for focusing gradient magnet (RGF) and defocusing gradient magnet (RGD), correct the chromaticities to (-10,-10).

## FEEDDOWN FROM END-SHIM

The orbit of a particle moving in the straight magnet (see Fig. 1) can be described by

$$x = x_o + \sqrt{R^2 - z^2} - R \approx x_o - d\left[\frac{2z}{L}\right]^2 \quad (2)$$

$$d = R(1-\cos(\theta/2)) = R\theta^2/8 = L\theta/8 \quad (3)$$

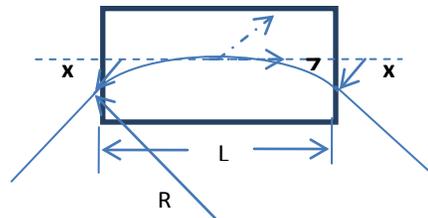

Figure 1: Orbit of the particle in a straight dipole

*Work supported by U.S. Department of Energy under contract No. DE-AC02-76CH03000.
#meiqin@fnal.gov

where *d* is called sagitta, and it was shown that the bend field feeddown from quadrupole componenets through the magnet can be compensated if the magnet is placed with the design orbit off by + (d/3) at the center. Then for the Recycler ring

$$x = -\frac{2}{3}d = -0.00781 \text{ m for RGF/RGD(ARC dipole)}$$

The magnetic field can be expressed by

$$B_y(x) = B_o(1 + b_1(x/r) + b_2(x/r)^2 + b_3(x/r)^3 + \cdots) \quad (4)$$

where the b1, b2 and b3 are the quadrupole, sextupole and octupole terms, r=1" is the radius at which the harmonic is measured, and the x is the offset through the multipoles. Then the contribution to the quadrupole of the end-shim are from 3 parts:

$b_1*(B_o/r)$ : quadrupole from end shim itself (in the order of $10^{-4}$)

$b_2*(2xB_o/r^2)$: feeddown effects from sextupoles at the closed orbit offset at the end-shim location (in the order of $10^{-4}$)

$b_3*(3x^2B_o/r^3)$: is feeddown effects from sextupole components, which is feeddowned from octupole components, at the closed orbit offset at the end-shim location (in the order of $10^{-6}$)

## RR-30 STRAIGHT SECTION

We replaced the RR-30 straight section with 8 "D-D half FODO cells", starts at 301 and ends at 309. We use 2 quads in a half-cell, with standard Recycler permanent focusing and defocusing quadrupoles. There are also 9 families of total 18 trim quads were installed in this section as second trombone insertion. We obtain the lattice of 30 straight section, as shown in Fig.2, with the maximum beta functions of 55 m. The trim quad currents are set to 0.

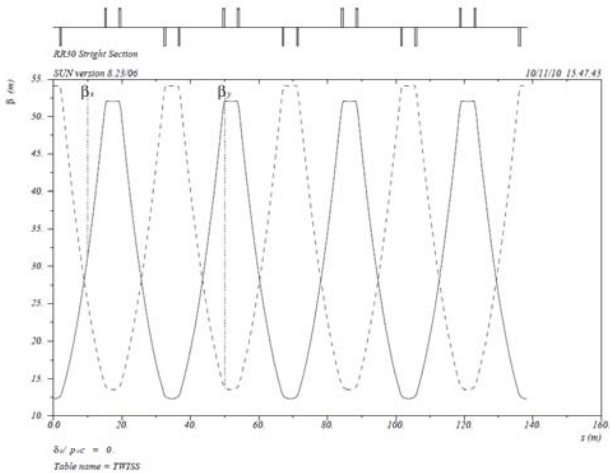

Figure 2: lattice functions of RR-30 FODO for Nova, each half cell has 2 permanent quads, with the standard permanent quads strength for focusing and defocusing.

## TUNING WITH END-SHIMS

The total tunes of the RR lattice with the RR-30 straight section shown in Fig. 2 are (25.0835, 24.2070), which differs from the working point tunes in (0.3415, 0.2080). Now only the quadrupoles of the "Standard end-shims" are used to tune the working point, with no current in all the trim quads in RR-30 and RR-60. We obtained the quadrupole component $(b_1)_F$ = -15.1786 and $(b_1)_D$= +5.6429 respectively for focusing gradient magnet (RGF) and defocusing gradient magnet (RGD) to get the tunes at $Q_x$=25.425, $Q_y$=24.415.

Table 1 lists a summary of the parameters of the end-shim for ARC gradient magnets. The beta-functions of the new Recycler Ring are shown in Fig. 3.

Table 1: End-shim parameters

| parameters | Results |
|---|---|
| chromaticities | (-10,-10) |
| Tunes | (25.425,24.415) |
| Arccell phase advance($\mu_x$,$\mu_y$) | (87.69672, 80.33256) |
| $(b_2)_{f(standard)}$ | -0.2318 |
| $(b_2)_{d(standard)}$ | -3.3985 |
| $(b_1)_d$ | +5.6429 |
| $(b_1)_f$ | -15.1786 |

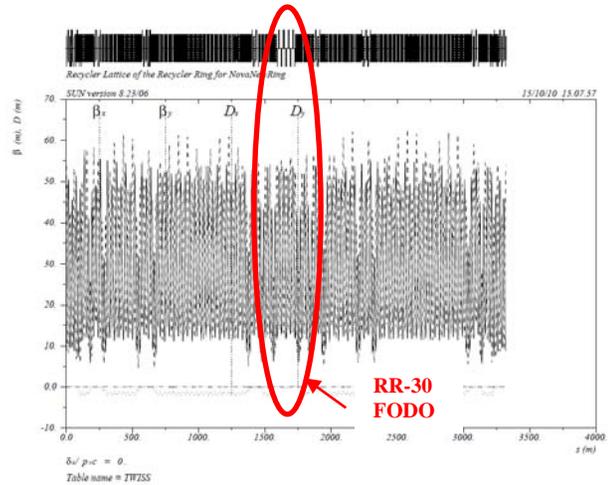

Figure 3: The Recycler lattice for Nova. The tunes are (25.425,24.415), Chromaticities are (-10,-10).

## DYNAMIC APERTURE OF THE REYCLER LATTICE FOR NOVA

We did the dynamic aperture (DA) tracking for the RR NOvA lattice to make sure that we still have large enough dynamic aperture for this lattice due to largely increased sextupoles of the end shims, as well as even more chromaticitie jump to (-20,-20) using 2 families of power sextupoles during the process of slip-stacking.

The scenario of beam injection into the Recycler Ring for NOvA [2] is shown in Fig. 4 There are two operating models, the first one shows 12 Booster batches, using slip stacking, last six Booster batchers are injected and will be slip stacked while the first six are circulating, the whole circulating time is 0.852 second. The second one shows a little longer time, 0.919 second, since there would be another 3 Booster batches injected and extracted for Project Mu2e [3] before last 6 Booster batches injection and slip stacking. The first 6 Booster batches would be circulating about 0.5 second which is about 45,000 turns (Recycler Revolution period is 11.12 ms), before the chromaticities are jumped up to (-20,-20) for slip stacking. Therefore, the particles only experience higher chromaticities for another 0.5 s (45,000 turns). The additional 10 unit of the chromaticities were obtained by two powered sextupole families in the Recycler ring. The estimate 95% normalized emittances is 14-16 $\pi$ mm.mrad, but we will use a conservative number of 18 $\pi$ mm.mrad for dynamic aperture tracking.

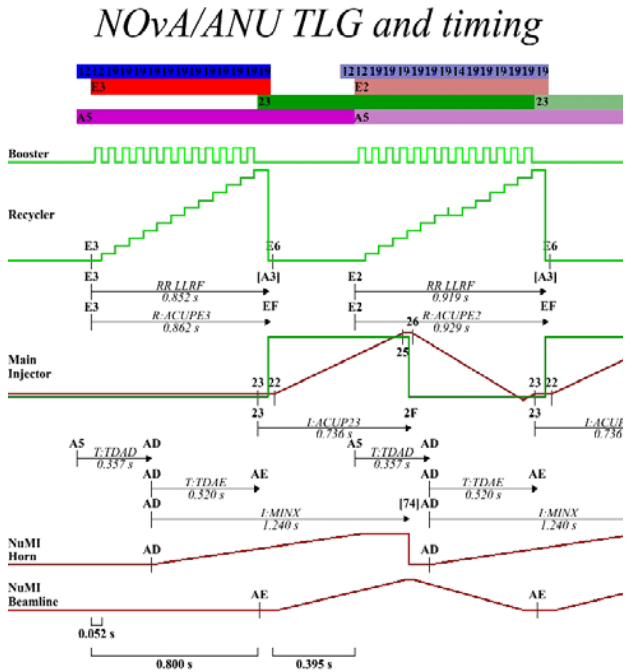

Figure 4: Nova/ANU TLG and timing. (Courtesy of Phil Adamson)

We did dynamic aperture tracking in Year 2000 [4] for the Recycler e-cool lattice. This time we used the same method described in Ref. [4]. Table 2 lists the results for different scenarios.

It was found that at normal operation mode (the chromaticities of (-2,-2)), we have average DA of 18 $\sigma_r$ in radius ($\sigma_r = \sqrt{\sigma_x^2 + \sigma_y^2}$). The beta-functions at MRK605 (starting point for tracking) in the RR lattice are $\beta_x$=8.838 m, $\beta_y$=53.820 m. the beam sizes with the emittance of 18 $\pi$mm.mrad is then $\sigma_x$=1.674 mm, $\sigma_y$=4.128 mm. The vacuum tube in RR-60 straight section is 3" round, 38.1 mm in both horizontal and vertical planes. Therefore, the physical aperture at MRK605 is (22.76 $\sigma_x$, 9.23 $\sigma_y$). From Table 2, we found that dynamic aperture is just about the same as the physical aperture at the scenario 3 and 4: for the particles with momentum deviation of (±2.0e-3) at chromaticities of (-20,-20). If the emittance coming from Booster can keep to ~14 $\pi$mm.mrad, this would not be a concern.

Table 2. lists average dynamic aperture (DA) at different conditions for Nova

| chromaticities | Number of turns Equivalent time | Momentum deviations | Average DA ($\sigma x, \sigma y$) |
|---|---|---|---|
| (-10,-10) | 50,000, 0.5s | 0 | 10.26 |
| (-20,-20) | 100,000, 1.0s | 0 | 10.15 |
| (-20,-20) | 50,000, 0.5s | 2.0e-3 | 9.82 |
| (-20,-20) | 50,000, 0.5s | -2.0e-3 | 9.35 |
| (-2,-2) | 100,000 | 0 | 18 |

## CONCLUSION

Newly designed "standard end-shims" can correct the chromicities to (-10,-10), and the tunes to the designed working point of (25.425,24.415). RR-30 straight section is now a 8 half-cell FODO lattice, using Fermilab standard permanent quads, with the maximum beta-functions less than 55 m. Second trombone insertion is also installed in RR-30 straight section for more tuning capability.

## REFERENCES

[1] D. E. Johson et al., "Corrections to the Fermilab Recycler Focusing with End-shim Changes," PAC'01, Chicago, June 2001, p. 2575 (2001); http://www.JACoW.org

[2] P. Adamson, "NOvA/ANU TLG and Timing," presentation on department meeting on October 13, 2010;

[3] M. Xiao, "Transport from the Recycler Ring to Anti-proton source Beam lines", this conference, TUPPR086;

[4] M. Xiao and T. Sen, "Dynamic Aperture Tracking For Fermilab Recycler Ring," PAC'01, Chicago, June 2001, p. 1717 (2001); http://www.JACoW.org